\documentclass[aps,preprint,showpacs,superscriptaddress,groupedaddress]{revtex4}  
\usepackage{graphicx}  
\usepackage{dcolumn}   
\usepackage{bm}        
\usepackage{amssymb}   
\usepackage{amsmath}
\usepackage{color}
\usepackage[normalem]{ulem}

\begin{document}

\title{Similarities and Differences Between Nonequilibrium Steady States and Time-Periodic Driving in Diffusive Systems}
\author{D. M. Busiello}
\affiliation{Dipartimento di Fisica `G. Galilei', Universit\'a di Padova, Via Marzolo 8, 35131 Padova, Italy}
\affiliation{Institute for Physical Science and Technology, University of Maryland, College Park, Maryland 20742, USA}
\author{C. Jarzynski}
\affiliation{Institute for Physical Science and Technology, University of Maryland, College Park, Maryland 20742, USA}
\affiliation{Department of Chemistry and Biochemistry, University of Maryland, College Park, Maryland 20742, USA}
\affiliation{Department of Physics, University of Maryland, College Park, Maryland 20742, USA}
\author{O. Raz}
\affiliation{Department of Chemistry and Biochemistry, University of Maryland, College Park, Maryland 20742, USA}
\affiliation{Department of Physics of Complex Systems, Weizmann Institute of Science, Rehovot, 76100, Israel}
\date{\today}

\begin{abstract}
A system that violates detailed balance evolves asymptotically into a nonequilibrium steady state with non-vanishing currents. Analogously, when detailed balance holds at any instant of time but the system is driven through  time-periodic variations of external parameters, it evolves toward a time-periodic state, which can also support non-vanishing currents.  In both cases the maintenance of currents throughout the system incurs a cost in terms of entropy production.  Here we compare these two scenarios for one dimensional  diffusive systems with periodic boundary condition, a framework commonly used to model biological and artificial molecular machines. We first show that the entropy production rate in a periodically driven system is necessarily greater than that in a stationary system without detailed balance, when both are described by the same (time-averaged) current and  probability distribution. Next, we show how to construct both a non-equilibrium steady state and a periodic driving that support a given time averaged probability distribution and current. Lastly, we show that although the entropy production rate of a periodically driven system is higher than that of an equivalent steady state, the difference between the two entropy production rates can be tuned to be arbitrarily small.  
\end{abstract}

\maketitle

\section{Introduction}

A system that is coupled to a thermal environment generically relaxes to an equilibrium state, in which its interesting properties can be calculated using the standard tools of statistical mechanics and thermodynamics. A similar unifying theory for all non-equilibrium phenomena is still lacking, although systems out of equilibrium have been investigated from various broad perspectives, including linear response theory \cite{kubo1957statistical,zwanzig2001nonequilibrium}, relaxation towards equilibrium \cite{Relaxation_montroll1957studies},  fluctuation theorems \cite{seifert_review_2012,Jarzynski2011},  nonequilibrium steady states (NESS) \cite{NESS_review_1} and systems with time-periodic driving \cite{Adiabatic_Molecular_motors_2007,Kolomeisky2007,Barato_PhysRevE.96.052120}. Many interesting results have been established within each of these frameworks, but much remains to be understood about the similarities and the differences between them.   

Systems that are constantly maintained away from equilibrium are of particular interest in biology and nano-science. There are two common ways to maintain a system out of equilibrium for arbitrarily long times: in the first, the system of interest is coupled to multiple environments, e.g.\ baths with different equilibrium properties such as temperature, chemical potential or voltage. In such cases, the constant fluxes between the baths drive the system into a steady state that is out of equilibrium, as it can only be maintained at the cost of thermodynamic resources (heat, fuel, photons, etc.) provided by the baths. These steady states are commonly referred to as Non-Equilibrium Steady States (NESS), and they are used to model a variety of biological processes, from photosynthesis \cite{Photosyntheses} in which photons are consumed in the carbon fixation process, through the synthesis of $ATP$ by an $ATP$-synthase where the chemical potential difference of $H^+$ ions across membrane is used to convert $ADP+P_i$ into $ATP$ \cite{Yasuda1998,Gaspard2007}, to molecular motors as kinesin \cite{Fisher2001} that consume $ATP$ molecules and generate directed motion for the transport of molecular cargo. 

An alternative method to maintain a system out of equilibrium is to vary, periodically with time, one or more parameters of the system, environment or the coupling between them. This type of driving is often referred to as stochastic pumping or thermal ratcheting. Stochastic Pumps (SP) provide simple models of classical and quantum heat engines \cite{gingrich2014efficiency,PhysRevX_Seifert_2015,Uzdin_PhysRevX.5.031044} or of the driving mechanism in artificial molecular motors \cite{Adiabatic_Molecular_motors_2007,Artificial_molecular_motors,Experimental_Engine,Kolomeisky2007,Leigh2003}, where periodic changes in macroscopic parameters such as temperature, pressure and pH keep the motor operating. 

Both NESS and SP are characterized by the existence of non-vanishing currents, a non-vanishing entropy production in the environment and a non-equilibrium probability distribution. It is therefore natural and potentially fruitful to ask: are SP and NESS essentially equivalent in terms of currents, probabilities and entropy production? In other words -- can any current, probability distribution and entropy production achievable using one type of driving can also be achieved with the other type as well? In terms of potential applications, this question can be stated as follows: can an artificial molecular motor driven by periodic changes in the environment exactly mimic a biological molecular motor driven by consuming fuel? For finite-state systems, this question has been recently addressed in \cite{Mimicking_PhysRevX}, where it was shown that SP and NESS are equivalent -- both systems can in principle have the same time-averaged probabilities, currents and entropy production rates. Interestingly, however, they are not equivalent in terms of fluctuations \cite{Rotskoff_PhysRevE.95.030101}: to match the current fluctuations of a NESS, SP must have a higher entropy production. 

In this manuscript we extend the NESS-SP comparison to overdamped systems that evolve diffusively in one dimension, whose dynamics are described by a Fokker-Planck equation on a ring. For artificial molecular motors, this model is typically more accurate than the discrete state case, which can be viewed as a coarse-grained version of a diffusive system. In the context of ``no pumping theorems'', a similar extension from discrete state models \cite{No-pump_mandal2014unification,No_pump_maes_2010,Rahav_No_Pump_with_interactions,chernyak_no_pump_2008,mandal2011No_pump_graphical,rahav2011extracting,rahav2008directed} to diffusive systems \cite{Horowitz2009} revealed a complete analogy between the two models.

As we show below, in the context of controllability diffusive systems are quite different from the discrete systems studied in \cite{Mimicking_PhysRevX}.  In discrete systems one can achieve full control of the system in the following sense: given a desired set of currents, entropy production, and probability distribution (which are time-independent in the case of NESS, or time-averaged over one period of driving in the case of SP), one can determine the parameters of the model required to achieve these targets.
By contrast, in diffusive systems full control of averaged currents and probability is possible, but these set a minimal bound on the corresponding entropy production rate, or even uniquely determine it for a NESS. Moreover, diffusive SP always generate more entropy production than NESS, when both drive the same averaged current and probability distribution. This suggests a natural optimization problem: finding the SP that achieves a target current and probability, with minimal averaged entropy production rate.

This manuscript is organized as follows: in Sec.~\ref{Sec:MathFramework} we introduce the mathematical framework to model diffusive SP and NESS systems. In Sec.~\ref{Sec:EntProdInEq} the entropy production inequality is derived. In Sec.~\ref{Sec:SP2NESS} we show full controllability of NESS in terms of current and probability distribution. This is done constructively, by obtaining the potential and velocity that generate a given target current and probability distribution. In Sec.~\ref{Sec:MimicNESSwithSP} we solve the analogous problem for SP; our construction requires several preliminary steps that illustrate crucial points of the analysis. In Sec.~\ref{Sec:Optimal_Protocol} we consider the optimization problem of minimizing the entropy production for a given target current and probability distribution. The general case of this problem is discussed in the appendix, App.(\ref{App:A}). We conclude in Sec.~\ref{Sec:Discussion} with discussions and proposals for further investigations.

\section{Mathematical framework}\label{Sec:MathFramework}

We aim to compare two types of driving in diffusive systems: the first is performed by the breakage of detailed balance in a time-independent system, and the second concerns the time-periodic variations of parameters of a detailed balanced system. To this end, let us first consider a diffusion process on a ring for which detailed balance holds instantaneously. The state of the system at time $t$ is denoted by $x(t) \in [0,1] $, using units such that the length along the ring is 1, and we identify $x=1$ with $x=0$ (periodic boundary conditions). The time-dependent probability density $P(x,t)$ obeys the Fokker-Planck equation
\begin{eqnarray}\label{Eq:FokkerPlanck_PD}
\partial_t P &=&  \gamma^{-1}\partial_x \Big[\left( \partial_x U \right) P \Big]  + D\partial_{xx}P,
\end{eqnarray}
where $U(x,t)$ is the time-periodic potential in which the system diffuses, and $\gamma$ and $D$ are the damping coefficient and diffusion constant, respectively. Motivated by the modeling of molecular motors, we assume that the diffusion constant $D$ does not depend on position, $x$. We also assume that $\gamma$ and $D$ satisfy the fluctuation-dissipation relation $\beta D = 1/\gamma$, where $\beta$ is the inverse temperature.

For the system to satisfy the detailed balance condition at all times, the potential must be periodic in $x$, namely $U(0,t)=U(1,t)$ for each $t$. Indeed, if the potential were suddenly frozen, the system would relax to an equilibrium state described by the Boltzmann distribution, with vanishing probability currents.  We denote the period of the driving by $T$, i.e. $U(x,t)=U(x,t+T)$. Eq.(\ref{Eq:FokkerPlanck_PD}) sets the basic model for a diffusive system driven by periodic variations of external parameters, commonly referred to as a \emph{stochastic pump} or as a \emph{thermal ratchet} \cite{Parrondo1998,Ratchets_PhysRevLett.71.1477}. In this model, the time dependence of the driving is encoded in the temporal variations of the potential $U(x,t)$. By Floquet theory, the probability distribution $P(x,t)$ of such a system converges with time to a unique solution that is periodic in both $x$ and $t$. We denote this periodic solution by $P^{ps}(x,t)$. 

A probability distribution $P(x,t)$ evolving under Eq.(\ref{Eq:FokkerPlanck_PD}) can be associated to a probability current
\begin{equation}
J(x,t) = - D \left[\partial_x P(x,t) + \beta P(x,t) \partial_x U(x,t) \right],
\label{Eq:J_DB}
\end{equation}
such that the probability obeys a continuity equation,
\begin{eqnarray}\label{Eq:Prob_Continuity}
\partial_t P(x,t)+\partial_x J(x,t)=0.
\end{eqnarray} 
The current associated with the periodic solution is
\begin{equation}
J^{ps}(x,t) = - D \left[\partial_x P^{ps}(x,t) + \beta P^{ps}(x,t) \partial_x U(x,t) \right].
\label{Eq:J_PS}
\end{equation}
Integrating both sides of Eq.(\ref{Eq:Prob_Continuity}) over one period of driving, at fixed $x$, gives
\begin{eqnarray}
P^{ps}(x,T)-P^{ps}(x,0) + T\, \partial_x \overline{J^{ps}(x,t)}=0.
\end{eqnarray} 
where the overbar denotes temporal averaging over one cycle. By the temporal periodicity of $P^{ps}$, the first two terms cancel out, hence $\overline{J^{ps}(x,t)}$ must be independent of $x$. This agrees with the intuitive expectation that, over one cycle, the same total probability flux flows across any point $x$. 

In addition to the probability densities and currents which we consider as the desired outcome of the driving, we are interested in the cost of the driving, given by the environment's entropy production rate \cite{seifert2005entropy},
\begin{equation}
\dot{S}^{ps}(t) = \int_0^1 dx \frac{J^{ps}(x,t)^2}{D P^{ps}(x,t)}.
\label{SdotPS}
\end{equation}

In order to compare the time-periodic scenario described above with time independent systems driven by an external force that violates detailed balance, we now introduce a description for the latter. Let us consider the  Fokker-Planck equation
\begin{eqnarray}\label{Eq:FP_SS}
\partial_t P &=&  \gamma^{-1}\partial_x \Big[\left( \partial_x U(x) \right) P \Big]  + D\partial_{xx}P - v \partial_x P
\end{eqnarray}
where $U(x)$ is spatially periodic, $U(0)=U(1)$, as before, but time-independent.
The term $v \partial_x P$ violates the detailed balance condition for any $v\neq 0$. The constant $v$ can be interpreted as a characteristic velocity of the probability flow, or alternatively as arising from an additional linear potential that breaks the spatial periodicity of $U(x)$, generating a non-conservative force.

For any $P(x,t)$ evolving under Eq.(\ref{Eq:FP_SS}) the instantaneous probability current is given by
\begin{equation}
J(x,t) = - D \left[\partial_x P + \beta P \partial_x \left(U(x) - \frac{v}{\beta D}x \right) \right]
\label{Eq:J_NDB}
\end{equation}
such that $J$ and $P$ satisfy the continuity equation Eq.(\ref{Eq:Prob_Continuity}). Note that Eq.(\ref{Eq:J_NDB}) reduces to Eq.(\ref{Eq:J_DB}) when $v=0$. 

For finite $v$ and bounded $U(x)$,  Eq.(\ref{Eq:FP_SS}) has a unique steady state solution, denoted by $P^{ss}(x)$, with an associated probability current
\begin{equation}
\begin{split}
J^{ss} &= - D \left[\partial_x P^{ss} + \beta P^{ss} \partial_x \left(U(x) - \frac{v}{\beta D}x \right) \right] \\
&= -D e^{-\beta U} \partial_x \left( e^{+\beta U} P^{ss} \right) + v P^{ss}
\label{Eq:J_SS}
\end{split}
\end{equation}
which is independent of $x$ since $\partial_x J^{ss} = -\partial_t P^{ss} = 0$. 

The entropy production rate of a NESS is given by an expression similar to Eq. (\ref{SdotPS}), which simplifies because $J^{ss}$ does not depend on $x$:
\begin{equation}
\dot{S}^{ss} = \left( J^{ss} \right)^2 \int_0^1  \frac{dx}{D P^{ss}(x)}.
\label{SdotSS}
\end{equation}

Our main interest in what follows is the controllability of NESS and SP in terms of probability distribution, current and entropy production. 
As we have just shown, in contrast with discrete state models, for diffusive systems in NESS the current and probability distribution uniquely define the entropy production, Eq.(\ref{SdotSS}).
We next establish that if a given NESS and SP support the same probability and current (after time-averaging in the case of the SP), then the entropy production in the SP is no less than that in the NESS.
This implies a lower bound, Eq.~\ref{eq:SdotPSbound}, for the time-averaged entropy production of a SP.

\section{Entropy Production Inequality}\label{Sec:EntProdInEq}
To show that the entropic cost of a SP is at least as high as that of a NESS supporting the same averaged current and probability distribution, we first note that given $J^{ss}(x)$, $P^{ss}(x)$, $J^{ps}(x,t)$ and $P^{ps}(x,t)$ the values of the entropy production rates $\dot S^{ss}$ and $\dot S^{ps}(t)$ are fully determined by Eqs.(\ref{SdotPS},\ref{SdotSS}). Therefore, in diffusive systems with a uniform diffusion constant we cannot impose the entropy production as an independent condition, as was done for discrete states systems \cite{Mimicking_PhysRevX}. This constitutes a fundamental difference between continuous and discrete-state systems: there is a minimal cost, in terms of entropy production, for driving a current through a diffusive system, whereas in discrete-state systems currents can have arbitrary small cost. We note that if the diffusion constant $D$ can be varied as a function of position and time, then the analysis in \cite{Horowitz2009} implies that diffusive systems would have the same behavior as discrete state systems. However, in contrast to the effective potential $\beta U(x,t)$ which can be manipulated by macroscopic parameters as light, concentrations, pH, temperature etc, manipulating the diffusion constant at the nano-scale is experimentally challenging, therefore we limit our discussion to systems in which it is constant.

Let us suppose, for the moment, full controllability of both NESS and SP in terms of their currents and probability distribution. Under this assumption, we can compare the entropy production of the two different scenarios, both supporting the same current and probability distribution. To this aim, consider the integral
\begin{equation}\label{Eq:I_def}
\mathcal{I} =  \int\frac{dt}{D T} \int dx \left[ \frac{J^{ps}(x,t)}{P^{ps}(x,t)} - \frac{J^{ss}}{P^{ss}(x)}\right]^2 P^{ps}(x,t) \geq 0.
\end{equation} 
Expanding the square in the integrand and rewriting each term, using  Eqs.(\ref{SdotPS},\ref{SdotSS}) along with simple manipulations, the following inequality can be derived:
\begin{equation}\label{Eq:Entropy_Inequality}
\mathcal{I} = \frac{1}{T} \int dt \dot{S}^{ps}(x,t) - \dot{S}^{ss} = \overline{\dot{S}^{ps}(t)} - \dot{S}^{ss} \geq 0
\end{equation}
Thus the entropy production rate of a NESS supporting a given steady state current $J^{ss}$ and probability distribution $P^{ss}(x)$ sets a lower bound on the average entropy production rate of a SP supporting the same (after time-averaging) current $\overline{J^{ps}}$ and probability distribution $\overline{P^{ps}(x)}$.
Using Eq.(\ref{SdotSS}) we obtain, explicitly,
\begin{equation}
\label{eq:SdotPSbound}
\overline{\dot{S}^{ps}(t)} \ge \overline{J^{ps}}^2 \int_0^1  \frac{dx}{D \overline{P^{ps}(x)}}.
\end{equation}
In what follows we will show that for any non-singular $P^{ps}(x,t)$, Eq(\ref{Eq:Entropy_Inequality}) is a strict inequality. However, the entropy production under periodic driving can be arbitrarily close to the bound set by the NESS. 

\section{Non-equilibrium Steady State}\label{Sec:SP2NESS}
\subsection{Current and Probability Distribution Controllability}
We first show that NESS can support any target probability distribution $P(x)$ and current $J$ as its steady state values. To this end, we aim at finding the velocity $v$ and potential $U(x)$ for which $P^{ss}(x)$ and $J^{ss}$, defined in Eq.(\ref{Eq:J_SS}),  are equal to the target values. This is achieved by inverting  Eq.(\ref{Eq:J_SS}), which can be viewed as a linear equation for $U(x)$. This gives, up to an additive constant:
\begin{equation}
U(x) = - \frac{J^{ss}}{\beta D} \int^x \frac{dy}{{P^{ss}(y)}}
- \beta^{-1} \log{P^{ss}(x)} + \frac{v x}{\beta D}.
\end{equation}
To determine $v$ we impose periodicity on $U(x)$. Using the periodicity of $P^{ss}(x)$ this gives:
\begin{equation}
v = J^{ss} \int_{0}^1 \frac{dy}{{P^{ss}(y)}}.
\end{equation}
These equations show how to build a NESS with desired $P^{ss}(x)$ and $J^{ss}$.

\subsection{Minimal entropy production in NESS}\label{Sec:MinEntNess}
As we have just seen, the steady state current and probability distribution of a NESS can be chosen independently -- the value of one of them does not constrain the value of the other. It is therefore natural to ask: given $J^{ss}$, what choice of probability distribution $P^{ss}$ minimizes the entropy production? The dual question, namely given $P^{ss}$ what $J^{ss}$ minimizes the entropy production, is trivial: when $J^{ss}=0$ (equilibrium conditions) there is no entropy production. The above question can be written as a simple minimization problem:
\begin{eqnarray}
\min_{P^{ss}(x)} \frac{(J^{ss})^2}{D}\int_{0}^{1}\left[ \frac{1}{P^{ss}(x)} + \lambda P^{ss}(x)  \right] dx
\end{eqnarray}
where $\lambda$ is a Lagrange multiplier associated with the normalization of $P^{ss}(x)$. In principle, there is additional constraint in the problem, the positivity of $P^{ss}(x)$, however, this is a non-holonomic constraint, and as we next show the optimal solution satisfies this constraint without having to impose it. The Euler-Lagrange equation for the above optimization problem is given by:
\begin{eqnarray}
-\left({P^{ss}(x)}\right)^{-2} + \lambda = 0,
\end{eqnarray}
which combines with normalization to give $P^{ss}(x)=1$. Thus the minimal entropy production required to drive a steady current $J^{ss}$ is, by Eq.(\ref{SdotSS}),
\begin{equation}
\label{eq:Sdotmin}
\dot S_{min}^{ss} = \frac{(J^{ss})^2}{D} , 
\end{equation}
which is achieved with a flat potential $U(x)=0$ and $v =J^{ss}$.

\section{Current and Probability Controllability in Stochastic Pumps}\label{Sec:MimicNESSwithSP}
In the previous section we showed how to construct a NESS with a target current and probability distribution. This task was simple, since the NESS has an explicit solution for the current (Eq.\ref{Eq:J_NDB}) in terms of the potential $U(x)$, the steady state distribution $P^{ss}(x)$  and the parameters, $v, D$ and $\beta$. In the current section, we consider the problem of controlling the time-averaged current and probability distribution in a system driven by a time-periodic potential, in which detailed balance holds instantaneously. Unfortunately, there is no simple explicit solution for $P^{ps}(x,t)$ in terms of $U(x,t)$ which can be inverted as in Sec.~\ref{Sec:SP2NESS}. However, we have considerable freedom in choosing the potential $U(x,t)$: as shown below, there are many choices that result in the same averaged probability distribution and current. An example of such a protocol can be constructed once the constraints set by detailed balance are taken into account.

To frame this discussion, it will be useful to imagine that we have already constructed a NESS with a desired current $J^{ss}$ and probability distribution $P^{ss}(x)$, and we want to design a SP that matches these values after time averaging, i.e.\ we aim to satisfy the conditions
\begin{subequations}
\label{eq:match}
\begin{eqnarray}
\label{eq:Jps-match}
\overline{J^{ps}} &\equiv& \frac{1}{T} \int_0^T J^{ps}(x,t) \, dt = J^{ss} \\
\label{eq:Pps-match}
\overline{P^{ps}}(x) &\equiv& \frac{1}{T} \int_0^T P^{ps}(x,t) \, dt = P^{ss}(x)
\end{eqnarray}
\end{subequations}
(Recall from Sec.~\ref{Sec:MathFramework} that $\overline{J^{ps}}$ does not depend on $x$.)

\subsection{Current Loop}
In discrete systems, a useful constraint on periodic driving is set by the ``no current loops'' condition, which states that if a system satisfies the detailed balance condition at a given instant in time, then there can be no instantaneous current loops, regardless of the instantaneous probability distribution (see \cite{Mimicking_PhysRevX} for details). We now show that a similar constraint holds for 1D diffusive systems.

Given instantaneous values of $P(x,t)$ and $J(x,t)$, a simple condition shows whether or not detailed balance is satisfied. Consider the integral
\begin{equation}
\mathcal{J}(t)= \int_{0}^{1} \frac{J(x,t)}{P(x,t)} dx,
\label{J_def}
\end{equation}
which has a natural physical interpretation: writing the  current density as the probability density times a mean local velocity, $J(x,t) = P(x,t) u(x,t)$ \cite{seifert_review_2012}, $\mathcal{J}(t)$ is  the instantaneous spatial average of this local velocity \footnote{A similar idea was recently discussed in \cite{Zia2017nonequilibrium}.}.
Using the spatial periodicity of $P(x,t)$ and $U(x,t)$ along with Eq.(\ref{Eq:J_NDB}), we obtain
\begin{eqnarray}
\mathcal{J} =  -D\int_{0}^{1} \partial_x\left[\log P+\beta U-\frac{v}{  D}x\right]dx = v
\label{Eq:J/P}
\end{eqnarray}
hence detailed balance is satisfied if and only if $\mathcal{J}=0$.

For the periodically driven SP that we consider here, detailed balance is satisfied at all times by assumption, hence $\mathcal{J}(t)=0$. As $P^{ps}(x,t)$ is necessarily positive, this condition implies that $J^{ps}(x,t)$ changes its sign as a function of $x$ -- this is the no-current-loops condition analogous to the one derived in \cite{Mimicking_PhysRevX}. Thus we cannot satisfy Eq.~\ref{eq:Jps-match} by demanding that $J^{ps}(x,t) = J^{ss}$; rather, $J^{ps}(x,t)$ must depend non-trivially on both $x$ and $t$.

An additional consequence of the condition $\mathcal{J}(t)=0$ is that the entropy production inequality, Eq.(\ref{Eq:Entropy_Inequality}), is a strict inequality. By Eq.(\ref{Eq:I_def}), $\mathcal{I}=0$ only when 
\begin{eqnarray}\label{Eq:EntropyEqualityCondition}
\frac{J^{ps}(x,t)}{P^{ps}(x,t)} = \frac{J^{ss}}{P^{ss}(x)} 
\end{eqnarray}  
for all $x$ and $t$, which in turn implies that the sign of $J^{ps}(x,t)$ is the same as that of $J^{ss}$. This, however, is inconsistent with the requirement that $J^{ps}(x,t)$ change sign as a function of $x$.  We conclude that ${\mathcal I}>0$, hence $\overline{\dot{S}^{ps}} > \dot{S}^{ss}$.

Given an instantaneous probability distribution $P(x,t)$ and current density $J(x,t)$ satisfying ${\mathcal J}(t)=0$, we can use Eq.(\ref{Eq:J_DB}) to obtain, up to an additive constant,
\begin{equation}
\label{Eq:U_from_P_and_J}
U(x,t) = - \frac{1}{\beta D} \int^x \frac{J(y,t)}{P(y,t)} dy - \beta^{-1} \log P(x,t)
\end{equation}
which satisfies the condition $U(1,t) = U(0,t)$.
Eq.(\ref{Eq:U_from_P_and_J}) gives the time-dependent potential $U(x,t)$ that generates the current pattern $J(x,t)$ for the probability distribution $P(x,t)$.

\subsection{Compatibility of $P(x,t)$ with detailed balance}
So far we have discussed the constraint between $P(x,t)$ and $J(x,t)$ imposed by the condition of detailed balance, namely ${\cal J}(t)=0$, and we have shown how to construct $U(x,t)$ from $P(x,t)$ and $J(x,t)$, at any instant in time (Eq.~\ref{Eq:U_from_P_and_J}).
It is natural to ask next: given a smooth, normalized $P(x,t)$, does there always exist a time-dependent potential $U(x,t)$ such that $P(x,t)$ is a solution of Eq.~\ref{Eq:FokkerPlanck_PD}?
In other words, is any well-behaved $P(x,t)$ compatible with detailed balance?
Naively, one might expect the answer to be negative, as the detailed balance condition sets a constraint on $J(x,t)$ and therefore on the time derivative of $P(x,t)$. Fortunately, this is not the case.  It can be shown that an arbitrary well-behaved (smooth and normalized) $P(x,t)$ can be driven by a time dependent detailed balance periodic potential. 

To establish this result we first construct, given $P(x,t)$, the corresponding current $J(x,t)$ that is compatible with detailed balance. From the continuity equation (Eq.\ref{Eq:Prob_Continuity}) we have:
\begin{eqnarray}\label{Eq:J_of_P}
J(x,t) &=& J(0,t) - \int_{0}^{x}\partial_t P(x',t) dx'
\end{eqnarray}
which necessarily satisfies the periodicity condition $J(0,t)=J(1,t)$ as $\partial_t\int_{0}^{1} P(x',t) dx'=0$ for normalized probabilities.
Thus the continuity equation dictates $J(x,t)$ up to a time dependent function, $J(0,t)$.

Next, we impose the constraint of detailed balance, $\mathcal{J}(t)=0$. Substituting Eq.(\ref{Eq:J_of_P}) into Eq.(\ref{J_def}) gives:
\begin{eqnarray}
\mathcal{J}(t) = \int_{0}^{1}\frac{J(0,t) - \int_{0}^{x}\partial_t P(x',t) dx'}{P(x,t)}dx ,
\end{eqnarray}
and by setting the right side to zero we arrive at
\begin{eqnarray}\label{Eq:J(L)_from_P(t)}
J(0,t)=\frac{\int_{0}^{1}\frac{\int_{0}^{x}\partial_t P(x',t) dx'}{P(x,t)}dx} {\left(\int_{0}^{1}\frac{1 }{P(x,t)}dx\right)}.
\end{eqnarray}
In other words, under the assumption of detailed balance $P(x,t)$ uniquely determines $J(x,t)$, and then the two together determine the potential $U(x,t)$, by Eq.(\ref{Eq:U_from_P_and_J}).

The next challenge is to choose a periodic $P^{ps}(x,t)$ such that (i) its time average is equal to $P^{ss}(x)$ (Eq.(\ref{eq:Pps-match})), and (ii) the corresponding time-averaged current is equal to $J^{ss}$ (Eq.(\ref{eq:Jps-match})). An explicit construction with these properties is shown in the next subsection.  

\subsection{Constructing $U(x,t)$ to generate a desired $\overline{P^{ps}}(x)$ and $\overline{J^{ps}}$ }
We begin by defining a dimensionless time $s=t/T$, and we consider how both $\overline{P^{ps}}(x)$ and $\overline{J^{ps}}$ scale with the total period of cycling, $T$, for a given choice of $P^{ps}(x,s)$.
We obtain:
\begin{eqnarray}
\overline{P^{ps}}(x) &=& \int_0^1 P^{ps}(x,s)ds \\
\overline{J^{ps}} &=& \frac{1}{T} \int_0^1 J^{ps}(x,s) \, ds ,
\end{eqnarray}
using Eqs.(\ref{Eq:J_of_P}) and (\ref{Eq:J(L)_from_P(t)}) to construct $J^{ps}$ from $P^{ps}$.
Thus $\overline{P^{ps}}(x)$ does not vary with $T$, while $\overline{J^{ps}}$ scales as $1/T$. Similarly, the time reversal of  $P^{ps}(x,s)$ defined by $P^{tr}(x,s)=P^{ps}(x,1-s)$ has the same temporal average as $P^{ps}(x,s)$, but the corresponding averaged current has opposite sign: $\overline{P^{ps}}(x) = \overline{P^{tr}}(x)$ and $\overline{J^{ps}} = - \overline{J^{tr}}$. 
Therefore, to satisfy Eqs.(\ref{eq:Jps-match}) and (\ref{eq:Pps-match}), we can choose a probability distribution $P^{ps}(x,t)$ with the desired temporal average and with a non-vanishing averaged current, and then match the averaged current by the rescaling of $T$ and its sign by time reversal. 
Lastly, given $P^{ps}(x,t)$ and $J^{ps}(x,t)$, we can use Eq.(\ref{Eq:U_from_P_and_J}) to construct $U(x,t)$. 

Importantly, the construction above has a lot of freedom: the only constraints on  $P^{ps}(x,s)$ are its time average, positivity, smoothness and a non-vanishing average current. This freedom implies that there exist many periodic potentials generating the same time-averaged current and probability distribution. We now illustrate this procedure with a simple example. 

\subsubsection{An Example for A Protocol}
\begin{figure}
	\centering
	\includegraphics[width=0.5\linewidth]{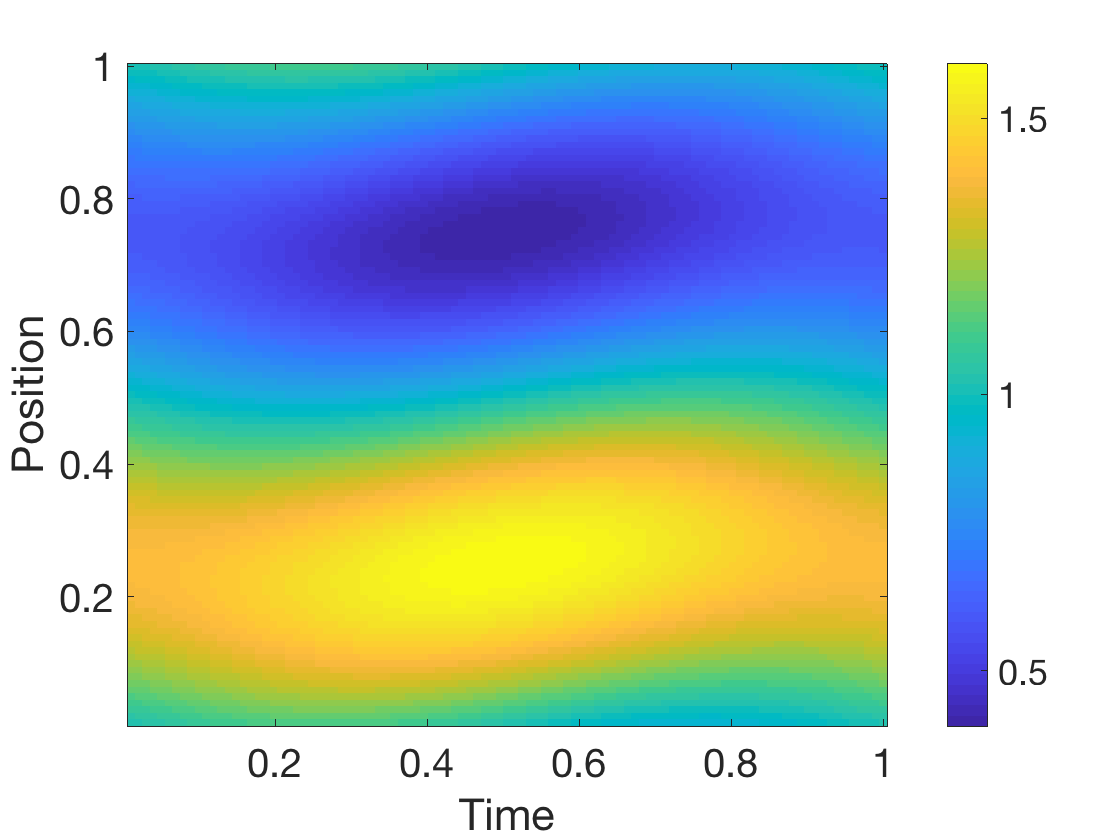}\\
	\includegraphics[width=0.5\linewidth]{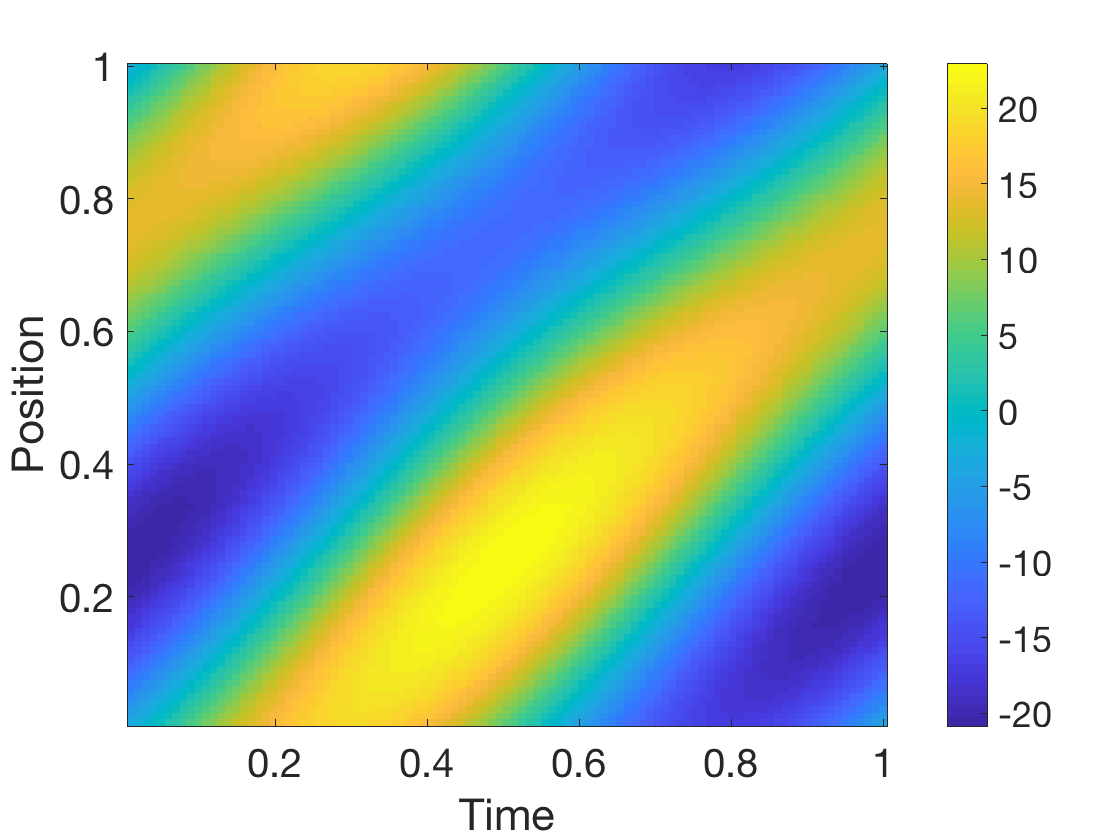}\\
	\includegraphics[width=0.5\linewidth]{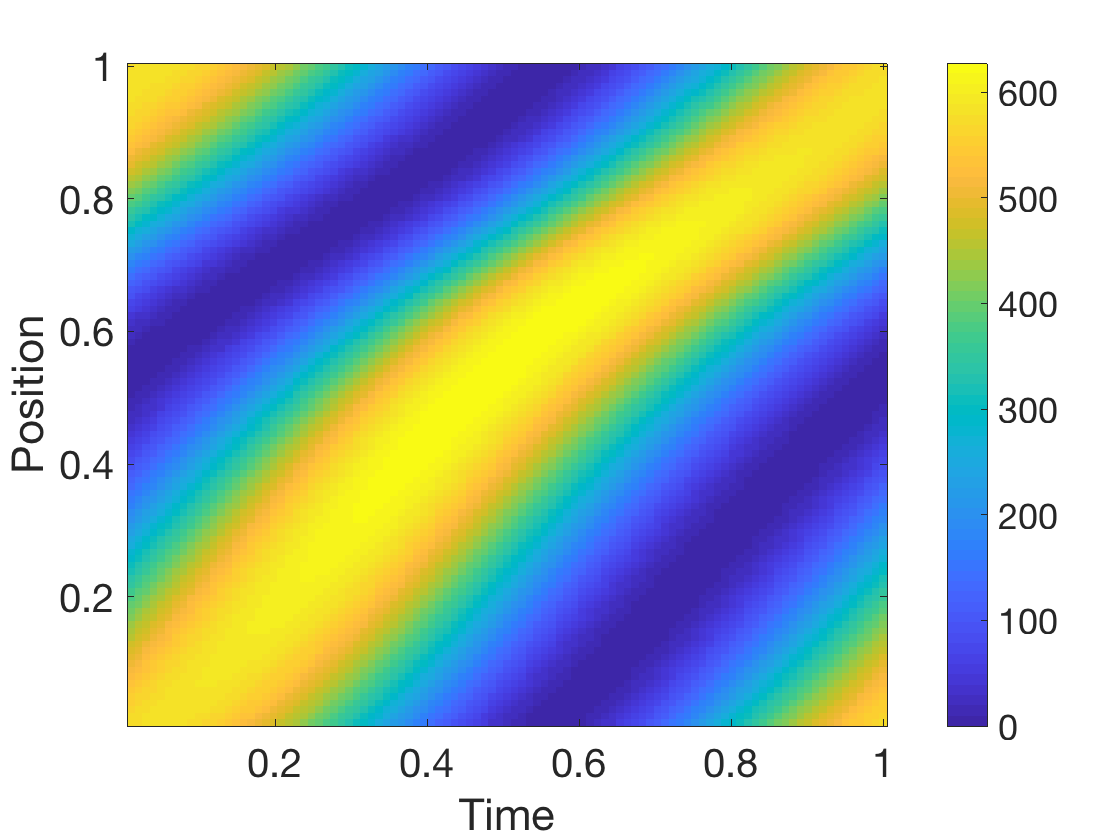}
	\caption{{\bf The example in Eq.(\ref{Eq:Example})}. Upper panel: $P(x,s)$, given in Eq.(\ref{Eq:Example}). Middle panel: the corresponding $J(x,s)$ for which the driving is assured to be detailed balance. Lower panel: the corresponding driving $U(x,u)$, given by Eq.(\ref{Eq:U_from_P_and_J})}
	\label{fig:pxtexample}
\end{figure}

Let us construct $U(x,t)$ that drives a time-averaged current and probability distribution
\begin{equation}
\overline{J^{ps}}=1 \quad,\quad
\overline{P^{ps}}(x) = 1 + 0.5\sin(2\pi x).
\end{equation}
As discussed above, $P^{ps}(x,s)$ can be chosen arbitrarily, provided it is positive, normalized and has the correct time average and non-vanishing current. The specific choice 
\begin{eqnarray}
P^{ps}(x,s) = 1 + 0.5\sin(2\pi x) + 0.1 \sin(2\pi( s - x )),\label{Eq:Example}
\end{eqnarray}
gives the desired time averaging $\overline{P^{ps}}(x)$.
Eq.(\ref{Eq:J_of_P}) implies in this case
\begin{equation}
J^{ps}(x,s) = \frac{1}{T} \Bigl( J^{ps}(0,s) + 0.1 \Bigl[ \sin(2\pi( s - x )) - \sin (2 \pi s) \Bigr] \Bigr)\nonumber
\end{equation}
where the expression for $J^{ps}(0,s)$, although analytical, is cumbersome and is not given explicitly. {To match the target $\overline{J^{ps}}$, we further set $T \approx 0.58$}.
Figure \ref{fig:pxtexample} shows $P^{ps}(x,s)$ for this example, as well as the corresponding  $J^{ps}(x,s)$ and $U(x,s)$; the latter was calculated numerically using Eq.(\ref{Eq:U_from_P_and_J}) with $\beta=1$ and $D=1$.

\section{Optimal driving Protocol} \label{Sec:Optimal_Protocol}
As we have seen, there is considerable freedom in constructing a protocol $U(x,t)$ that drives a target $\overline{P^{ps}}(x)$ and $\overline{J^{ps}}$. Moreover, in Section \ref{Sec:EntProdInEq} it was shown that the entropy production rate of a SP always exceeds that of a NESS, when both share the same time averaged probability distribution and current; see Eq.(\ref{Eq:Entropy_Inequality}).  It is therefore natural to look for the protocol $U(x,t)$ that drives the target averages at the minimal entropy production cost. In other words, we would like to solve the following minimization problem:
\begin{eqnarray}
\min_{U(x,t),\; T} \left[\frac{1}{T}\int_0^T \dot S^{ps}[U(x,t)]dt \right]
\end{eqnarray}  
under the constraints:
\begin{eqnarray}
\overline{P^{ps}[U(x,t)]}(x) &=& P^{target}(x)\\
\overline{J^{ps}[U(x,t)]} &=& J^{target}
\end{eqnarray}

Solving the optimization problem directly is challenging, but unnecessary: it is possible to construct  a specific protocol that asymptotically approaches the bound. The construction of this protocol for generic $\overline{P^{ps}}(x)$ is given in Appendix \ref{App:A}. In this section a simple example of this construction is demonstrated.  

Let us consider driving a given current, $\overline{J^{ps}} = J_0$, with $\overline{P^{ps}}=1$. This example is of special interest since for a current $J_0$, the bound on $\overline{\dot{S}^{ps}}$, given by $\dot S^{ss}$ of a NESS with the same averaged probability and current, is minimal for a uniform probability distribution $P^{ss}(x)=1$, as discussed in Section \ref{Sec:MinEntNess}. To construct the driving, we consider a probability distribution of the form $P(x,t) = f(x-ut)$ for a positive, normalized, non-uniform function $f(x)$. In other words, we consider a probability distribution with a fixed shape that moves at a constant velocity $u$. The cycle time for this driving is $T=u^{-1}$, and the spatial symmetry implies that the temporal average of $P^{ps}(x,t)$ is $\overline{P^{ps}}(x)=P^{ss}=1$. In this case, the continuity condition in Eq.(\ref{Eq:J_of_P}) implies that 
 \begin{eqnarray}
 J^{ps}(x,t) =J^{ps}(0,t) + u \Big( f(x-ut)-f(-ut)\Big) ,
 \end{eqnarray}
 where $J^{ps}(0,t)$ is set by the detailed balance condition $\mathcal{J}=0$, Eq.(\ref{Eq:J(L)_from_P(t)}), to be
 \begin{eqnarray}\label{Eq:alpha_def}
 J^{ps}(0,t) = -u\left(\alpha - f(-ut)\right)  \quad,\quad  \alpha \equiv \left[ \int_0^1 \frac{dx}{f(x)} \right]^{-1}  \in (0,1).
 \end{eqnarray}
 The averaged current is therefore given by
 \begin{eqnarray}
 \overline{J^{ps}} = u\left(1 - \alpha \right).
 \end{eqnarray} 
 The target current is set to be $\overline{J^{ps}} = J_0$, which gives us
 \begin{eqnarray}
 u = \frac{J_0}{1-\alpha}.
 \end{eqnarray} 
 Substituting the above results into Eq.(\ref{SdotPS}), we get after some trivial algebra:
 \begin{eqnarray}
 \dot{S^{ps}} &=& \frac{J_0^2}{D(1 - \alpha)},
 \end{eqnarray}
 which -- as one might have guessed by translation symmetry -- does not depend on time.
We see that $\dot S^{ps}$ is minimal when $\alpha^{-1}=\int_0^1f(x)^{-1}dx$ is maximal. But this integral is not bounded from above: for example, in the limit $P(x,t)\rightarrow \delta(x-ut)$ the integral $\int_0^1f(x)^{-1}dx$ diverges, and we then get $\alpha\rightarrow 0$ hence $\dot S^{ps} \rightarrow \overline{J^{ps}}^2/D$.
Comparing with Eq.(\ref{eq:Sdotmin}) we see that, in this limit, the entropy production of the periodically driven state approaches the bound set by the corresponding steady-state value.
 
\section{Discussion}\label{Sec:Discussion}
In this work we discussed similarities and differences between two types of driving that maintain a diffusive system on a ring out of equilibrium: periodic variations of a potential along the ring, and static driving by breaking the detailed balance condition. We have shown that the two scenarios can drive any averaged current and probability distribution, but in contrast to discrete state Markovian systems  there is no full control in terms of the averaged entropy production. Moreover, it was shown that the averaged entropy production of a steady-state driving is smaller than that of a system driven by periodic changes in the potential that achieves the same averaged current and probability distribution. In terms of applications, this implies that the common driving in biological molecular motors -- burning fuel and reaching a steady state -- is more efficient than the common driving of artificial molecular motors, namely periodic variation of external parameters. This result is different than what was obtained in a coarse-grained description of the same system -- discrete state Markovian modeling.
 
Many important aspects were not discussed in this work and they could be subjects to future investigations. These include  mapping between NESS and SP that matches other features (e.g. heat or work in heat engines \cite{Engines_PhysRevLett.116.160601}, current fluctuations \cite{Rotskoff_PhysRevE.95.030101} or entropy production fluctuations \cite{Barato_PhysRevE.96.052120}), as well as comparison of these two types of driving to other non-equilibrium scenarios. 

\section{Acknowledgments}
We thank R. Zia for pointing out the physical interpretation of $\mathcal{J}$. O.R. is supported by a research grant from Mr. and Mrs. Dan Kane  and the Abramson Family Center for Young Scientists.
C.J. acknowledges financial support from  the U.S. Army Research Office under contract number W911NF-13-1-0390.

\appendix
\section{Minimizing Entropy Production for non-uniform $\overline{P^{ps}}(x)$}\label{App:A}
In Sec.~\ref{Sec:Optimal_Protocol} we analyzed a specific example where the entropy production rate of a stochastic pump can get arbitrarily close to that of a NESS with the same time averaged current and probability distribution. In this appendix we generalize this construction for  arbitrary target $\overline{P^{ps}}(x)$ and $\overline{J^{ps}}$. 

Analogously to the construction in Sec.~\ref{Sec:Optimal_Protocol}, we choose the probability distribution to be a translating profile, $P^{ps}(x,t) = f(x-x_0(t))$, where $x_0(t)$ changes monotonically from $0$ to $1$ over the interval $0\le t \le T$.
We first show that 
by appropriately choosing $x_0(t)$ we can construct the target time averaged probability:
\begin{eqnarray}\label{Eq:dot_x_P_ps}
\overline{P^{ps}}(x) = \frac{1}{T}\int_0^T f(x-x_0(t))dt =  \frac{1}{T} \, f*\frac{1}{\dot x_0},
\end{eqnarray} 
where $f*(1/\dot x_0)$ denotes the convolution of the functions $f(x)$ and  $1/\dot x_0(x)$, and $\dot x_0(x)$ is the velocity $dx_0(t)/dt$ expressed as a function of position, $x_0(t)=x$.
To gain some intuition, consider the example $f=\delta(x-x_0(t))$. By controlling the speed at which this delta function moves across each point we can manipulate the time averaged probability at this point.
Specifically, for this example Eq.(\ref{Eq:dot_x_P_ps}) gives us $\dot x_0(x) = 1/\left[ T\overline{P^{ps}}(x) \right]$.
More generally, applying the convolution theorem of Fourier transforms to Eq.(\ref{Eq:dot_x_P_ps}), we obtain
\begin{equation}\label{Eq:A2}
\frac{1}{\dot x_0(x)} = T \sum_n e^{i2\pi nx} \, \frac{\overline{P^{ps}}_n}{f_n}
\end{equation}
where $\overline{P^{ps}}_n$ and $f_n$ are the $n$'th discrete Fourier coefficients of $\overline{P^{ps}}(x)$ and $f(x)$. Note that the above equation shows that not any $f(x)$ can serve for our construction -- for example, if the right hand side of the above equation vanishes for some $x$ then the corresponding $\dot x_0(x)$ diverges. This can be intuitively understood by considering the extreme scenario: if $f(x)=1$, then one cannot match any probability distribution by averaging over $f(x-x_0(t))$, namely over translated versions of $f(x)$. Nevertheless, given any $\overline{P^{ps}}(x)$, one can always choose appropriate $f(x)>0$ which is narrow enough such that the expression in the right hand side of Eq.(\ref{Eq:A2}) is strictly positive, as is evident from the delta-function example above.

From the function $\dot x_0(x)$, we construct $t(x_0) = \int_0^{x_0} dx / \dot x_0(x)$, and then invert $t(x_0)$ to obtain $x_0(t)$.

Next, let us consider the current. By Eq.(\ref{Eq:J_of_P}),
\begin{eqnarray}
\label{eq:Jpsxt}
J^{ps}(x,t) = J^{ps}(0,t) +\dot{x}_0(t)\Big[ f(x-x_0(t))-f(-x_0(t)) \Big]
\end{eqnarray}  
For the detailed balance condition to hold, Eq.(\ref{Eq:J(L)_from_P(t)}) implies that
\begin{eqnarray}
\label{eq:Jps0t}
J^{ps}(0,t) = \dot{x}_0(t)\left[ f(-x_0(t))-\alpha \right] \,
\end{eqnarray}
where, as in Eq.(\ref{Eq:alpha_def}), 
\begin{eqnarray}
\label{eq:alphadef}
\alpha \equiv \left[ \int_0^1 \frac{dx}{f(x)} \right]^{-1} .
\end{eqnarray}
Eqs.~\ref{eq:Jpsxt} and \ref{eq:Jps0t} then give us
\begin{eqnarray}
\label{eq:Jpsxt_final}
J^{ps}(x,t) =\dot{x}_0(t)\left[ f(x-x_0(t))-\alpha \right] .
\end{eqnarray}
Let us set the target time averaged current to be $\overline{J^{ps}}=J_0$ for arbitrary $J_0>0$. With this choice the cycle time $T$ solves the equation
\begin{eqnarray}
\label{eq:setJps}
J_0 = \frac{1}{T}\int_0^T J^{ps}(x,t)dt = \frac{1-\alpha}{T},
\end{eqnarray}
where in the last equality we changed the variable of integration from $t$ to $x_0$.

Lastly, substituting Eq.(\ref{eq:Jpsxt_final}) into Eq.(\ref{SdotPS}), it can be shown that the entropy production rate at each instant is given by
\begin{eqnarray}
\dot S^{ps}(t) = \frac{\dot x_0^2}{D} \left(1-\alpha\right)
\end{eqnarray}
The time averaged total entropy production is therefore given by
\begin{eqnarray}
\overline{\dot{S}^{ps}} = \frac{1-\alpha}{TD} \int_0^T \dot x_0^2dt = \frac{J_0}{D}\int_0^1 \dot x_0(x_0)dx_0 ,
\end{eqnarray}
using Eq.(\ref{eq:setJps}).
In the limit $f(x)\rightarrow \delta(t)$ Eqs.(\ref{Eq:dot_x_P_ps}) and (\ref{eq:alphadef}) give us
\begin{equation}
\dot x_0(x) \rightarrow \left[ T\overline{P^{ps}}(x)\right]^{-1}
\quad , \quad
\alpha \rightarrow 0
\end{equation} 
hence $J_0 \rightarrow 1/T$ and 
\begin{eqnarray}
\overline{\dot{S}^{ps}} \rightarrow \frac{J_0}{DT}\int_0^1 \left[\overline{P^{ps}}(x)\right]^{-1}dx
\rightarrow \frac{\overline{J^{ps}}^2}{D}\int_0^1 \left[\overline{P^{ps}}(x)\right]^{-1}dx
\end{eqnarray}
which  is the bound on the entropy production rate of periodic driving with the corresponding time averaged current and probability (Eq.(\ref{eq:SdotPSbound})).

\end{document}